**Using genome-wide expression compendia to study microorganisms**

Alexandra J. Lee[1], Taylor Reiter[2], Georgia Doing[3], Julia Oh[3], Deborah A. Hogan[4], Casey S. Greene[2,5]

[1] Genomics and Computational Biology Graduate Program, University of Pennsylvania, Philadelphia, PA, USA

[2] Department of Biochemistry and Molecular Genetics, University of Colorado School of Medicine, Denver, CO, USA

[3] The Jackson Laboratory for Genomic Medicine, Farmington, Connecticut, USA

[4] Department of Microbiology and Immunology, Geisel School of Medicine, Dartmouth, Hanover, NH, USA

[5] Department of Pharmacology, University of Colorado School of Medicine, Denver, CO, USA

**Abstract**

A gene expression compendium is a heterogeneous collection of gene expression experiments assembled from data collected for diverse purposes. The widely varied experimental conditions and genetic backgrounds across samples creates a tremendous opportunity for gaining a systems level understanding of the transcriptional responses that influence phenotypes. Variety in experimental design is particularly important for studying microbes, where the transcriptional responses integrate many signals and demonstrate plasticity across strains including response to what nutrients are available and what microbes are present. Advances in high-throughput measurement technology have made it feasible to construct compendia for many microbes. In this review we discuss how these compendia are constructed and analyzed to reveal transcriptional patterns.

**Introduction**

Genome-wide transcriptional profiling measures the expression of all genes within a given sample.[1,2] This profile captures a snapshot of an organism's cellular state – what genes are active and how much they change in response to an environmental condition[3] or stimulus[4]. Consequently, transcriptional patterns can reveal the biological processes and possible mechanisms that contribute to traits including virulence[5–7], antibiotic resistance[8–10], metabolic versatility[11,12] and adaption[13]. These traits are of interest because they pertain to anthropocentric processes like microbial infection and bioreactor design,[14,15] inform our understanding of biological mechanisms in multicellular eukaryotes,[16] and underlie ecological cycles of biotransformations.[17] Therefore, transcriptomic studies are commonly used to examine trait-associated genes and their regulation.

Early experiments revealed the importance of transcriptional regulation in microbes. For example, experiments in the model organisms *Escherichia coli* (*E. coli*) and *Saccharomyces cerevisiae* revealed that common gene expression responses were elicited by different environmental stressors.[18,19] Studies in *Pseudomonas aeruginosa* (*P. aeruginosa*), an opportunistic gram-negative pathogen, found transcription factors including the global regulator LasR control the expression of a number of extracellular factors that contribute to virulence[6,20,21] including proteases[5,22]. Overall, by studying the global transcriptome response, we can start to understand the mechanisms of traits of interest.

Microbial transcription in response to microbial interactions and environmental cues is complex. Genome organization affects transcription through diverse mechanisms, including factors like 3-dimensional organization[23], gene proximity[24], and promoter location[25]. Transcriptional regulators interact with internal and external cues to achieve transcription programs that reflects their environment. For example, in microbial quorum sensing (QS), a cell-cell communication process that allows microbes to respond to population density through signal molecules, microbes produce and respond to signals that facilitate adaptation to varying conditions.[26,27] QS regulators can also impact environmental responses by regulating other transcription factors such as the oxygen-sensitive Anr in *P. aeruginosa*[28]. Anr activity is higher in QS-defective strains that lack function of the LasR QS regulator. Thus, *lasR* mutants (LasR-), which are frequently isolated from CF patients, are more fit in microoxic conditions than their LasR+ counterparts.[29]

In addition to environmental cues, microbes tend to grow in polymicrobial communities where they sense and transcriptionally respond to other microbes. Both competitive and cooperative behaviors[30,31] influence phenotypes[32,33], eliciting interactions like the production of public goods, (cross-feeding)[34], resource consumption, interference competition[35], or coordinate production of phenotypes (increased virulence and antibiotic resistance[36]). For example, in co-infection of *S. aureus* and *P. aeruginosa*, *P. aeruginosa* exoproducts can select for *S. aureus* small colony variants that are aminoglycosides resistant.[36] Finally, these microbe-microbe interactions are also dependent on environmental factors. Doing *et al.* found that *P. aeruginosa* produced antifungal phenazines against *Candida albicans* (*C. albicans*), but that this antagonistic interaction depends on phosphate availability and *C. albicans* fermentation.[37,38] Even two different genotypes can influence each other as in citrate cross-feeding found by Mould *et al*.[39]

Given the context-specific nature of transcription, leveraging data across many experiments allows researchers to study how microbes regulate transcription of different genes and pathways across different conditions – to gain a more systems level understanding of the transcriptome. Gene expression compendia, which are integrated collections of experiments, are one solution for examining transcriptional patterns across contexts. In this review we describe how these compendia are constructed and the challenges faced as well as highlight analyses using compendia to reveal patterns of interest.

**Construction of microbial expression compendia**

For the purposes of this review, we defined an expression compendium to be a heterogeneous collection of more than 50 gene expression experiments assembled from data collected for diverse purposes. Notable existing microbial compendia can be found in Table 1. The construction of each of these compendia began with the collection of relevant gene expression experiments from public repositories like ArrayExpress[40], Gene Expression Omnibus (GEO)[41], Sequence Read Archive (SRA)[42] and others[43,44]. Experiments of interest were then downloaded from these public repositories. In the case of the compendia represented in Table 1, all experiments (i.e. samples deposited together) within a given compendia were measured on the same platform to avoid bias and maintain a uniform reference. Additional filtering of samples were optionally performed to ensure that removal of spurious random correlation between genes.[45] Next, the samples were normalized to allow for cross sample comparison. Filtering, consistency in platform and normalization ensure that the compendium data is uniformly processed, facilitating downstream cross-sample comparisons.

There are different normalization techniques available depending on the technology. As an example, the *P. aeruginosa* RNA-seq compendium started with expression profiles downloaded from SRA and then median-ratio (MR) normalized.[45,46] The authors evaluated well-known RNA-seq normalizations, transcripts per million (TPM) and trimmed mean of means (TMM)[47], which corrected for spurious correlations; however correlations between random pairs of genes were still elevated compared to using MR normalization, which was their preferred strategy. These RNA-seq normalization methods address systematic variation, including differences in library size (i.e. sequencing depth)[48] and gene length[49], allowing for between sample and gene comparisons. Similarly, there also exist systematic variation in measurements using array technology though the sources are different and include differences in preparation protocol (i.e., total quantity of starting RNA, dye labeling) or differences in processing (i.e., different scanners or runs). One of the well-established normalization methods for the Affymetrix GeneChip system, which most of the compendia in Table 1 used, is RMA[50] which is a quantile method. In comparison to other single label normalization methods, Bolstad *et al.*[51] reported that RMA successfully reduced bias at reasonable compute speed compared to other global normalization methods. A similar review of two-color array technology, performed by Yang *et al.*[52], showed that different global or location-based normalization methods should be performed depending on the set of control spots. In a couple cases, where the compendium integrated across different platforms, such as two different array technologies or combining array and RNA-seq, studies used quantile normalization.[53–55] Regardless of the technology used, expression levels between samples can vary due to technical reasons, mentioned above, and so it's important to use normalization methods to adjust for these differences in order to compare between two gene expression profiles for applications such as gene function prediction, transcription regulatory network (TRN) inference and feature extraction.

Most of the existing compendia in Table 1 did not apply batch correction. In one case, where the compendium combined array and RNA-seq data, ComBat[54] was applied. While normalization is necessary in the context of compendia and facilitates cross-sample comparisons, batch correction is an optional step, and its application depends on the experiments included in the compendia. See section 'Challenges integrating across experiments' for a discussion of batch correction.

As more transcriptome data are generated, repositories like refine.bio[55], COLOMBOS[56,57] PILGRM[58], and M[3D] [59] are being developed to provide easily downloadable compendia where the data has been uniformly processed for different bacterial species. In general, the abundance of data has facilitated the generation of compendia to study transcriptional patterns across experiments.

**Why use compendia: Benefits and applications of using compendia**

*Systems-level models*

The construction of compendia, which contain hundreds to thousands of samples, has opened the door to the development of computational approaches, especially machine learning methods that have been successful at prediction tasks[60] and pattern extraction[61] in computer science, to discover transcriptional patterns in microbes.

Compendia can contribute to helping us gain a systems-level understanding of microbial biology. One major goal for systems biology is to model how information is encoded, specifically to reverse engineer the hierarchy of the transcriptomic regulatory network (TRN).[62–68] Knowing the organization of a regulatory network allows us to control or optimize parts of the system, a necessary step for many biotechnological advances.[69–72] This task requires a large amount of heterogeneous data, which compendia provide, to identify shared patterns looking across a variety of interventions.[73]

Dimensionality reduction methods can also be deployed to extract key patterns in data and reveal the transcriptional relationships between sets of genes.[74] Applying dimensionality reduction models to compendia allows users to study changes in gene sets and reveal more subtle and possibly undiscovered signals that could be masked by strong signals (i.e. a large fraction of genes representing the same pathway).[75,76] For example, a denoising autoencoder trained on a *P. aeruginosa* compendium, ADAGE, captured regulation patterns and biological processes.[77] Tan *et al.* showed that co-operonic genes were weighted highly in the same latent variables and, similarly, KEGG gene sets were enriched in some latent variables. They also showed that function prediction using the ADAGE weight matrix was more accurate compared to using a randomly permuted gene weight matrix. Furthermore, the latent representation of the gene expression data detected existing subtle expression differences[77] and also revealed a new aspect of low phosphate response that depends on the media[78]. These latent variables were also shown to detect pathway-pathway relationships -i.e. pathways that co-occur in the same latent variable.[79] A similar dimensionality reduction analysis was performed applying a sparse autoencoder to a yeast compendium, where Chen *et al.* found latent variables represented pathways and other layers of biological abstractions.[80] In other studies, applying independent component analysis (ICA) to a compendium of transcriptome data revealed transcription modules.[64,68] Specifically, Rajput *et al.* identified differentially active modules that varied based on the conditions, which defined coordinated activity of modules that were involved in functions that influence *P. aeruginosa* pathogenesis.[68] These unsupervised approaches summarize patterns in the expression compendia that can abstract different layers of a biological system that are useful for understanding the interaction between different molecular processes as well as generating new hypothesis. Webtools were developed to facilitate the exploration of the summarized data, like ADAGE,[81] as well as to search through the experiments available in compendia such as the ones found in COLOMBOS[57,82,83], PILGRM[58] and others[84] in order to direct future research.

*Methodologies to leverage compendia*

With the breadth of transcriptional patterns captured by compendia, recent approaches have been developed that demonstrate how compendia can be used to put new experiments in the context of existing ones as well as to leverage the aggregation of patterns available to study genomic patterns. Lee *et al.* developed a general framework for distinguishing between common and experiment-specific differentially expressed genes, called SOPHIE (Specific cOntext Pattern Highlighting In Expression data).[85] This approach compares gene expression changes in their target experiment with changes in a background set of experiments thereby allowing researchers to interpret and prioritize patterns in differentially expressed genes. The authors demonstrated that SOPHIE successfully prioritized genes with small differences in expression that were directly due to the perturbation being studied and not due to condition-specific secondary effects. In general, reanalysis and mining of the experiments within these compendia can be facilitated by tools like SOPHIE[85] or algorithms like GAUGE[86], which automate sample group detection for downstream statistical analyses. Overall, approaches like SOPHIE can find patterns that generalize across compendia.

*Condition-specific responses*

Transcriptional profiling is a snapshot of an organism's state, which is a complex representation of the cellular state and functions. Understanding the information that is captured in these profiles is important, especially for microbes that sense and respond to their environment. For example, Kim *et al.* inferred *E. coli* cellular and environmental state, like growth phase or aerobic conditions, from a gene expression compendium and identified pathways that are associated with the genes that are most predictive of these cellular states.[54] In other examples, studies also used gene expression to annotate the functional roles of genes.[87,88] Overall, by using these compendia to make predictions we can learn what genes are involved in different environmental conditions or processes, which can improve our understanding of microbial condition-specific responses.

Importantly, the identification of conditional regulons requires the study of a response of interest across multiple conditions. The diversity of condition-specific responses has been elucidated in targeted studies that have examined expression profiles in response to multiple stimuli such as various stressors.[89] However, the comprehensive mapping of condition-specific responses is often beyond the scope of an individual experiment. The re-analysis and meta-analysis of publicly available data revealed subsets due to the natural differences in how separate groups studied related phenomena in a way that informed each other. For example, through compendium-wide analysis of the low phosphate response, Tan *et al*. identified a condition-specific element of the low phosphate signaling cascade.[78] This result would not have stood out from any individual experiment but was clear when the larger compendium was analyzed.

*Inspiration from non-microbial expression compendia*

Non-microbial gene expression compendia have also been generated and used for a variety of purposes, many of which may inspire future endeavors for microbial compendia.[90–96] A human-based gene, ortholog, or k-mer based tool could facilitate rapid searches of the microbial compendia to identify samples from different experiments with similar expression profiles. Transfer learning has also successfully transferred knowledge contained in publicly available data sets and databases to rare disease samples.[92,95] Such methods could be applied to better unravel pathway-level patterns for rare microbial species. Lastly, human compendia have been leveraged to identify alternative splicing[93], lessons which may be applied to the discovery of polycistronic transcripts directly from RNA-seq reads. Further research is needed to explore how lessons learned from human transcriptome compendia can best apply to microbial transcriptomics.

These studies demonstrate that the versatile data that is available in compendia provides a valuable resource to gain a systems level understanding of transcriptional signaling as well as to make predictions. Additionally a low dimensional representation of compendia capture transcriptional patterns that can reveal coordinated activity of gene sets and pathways as well as allows researchers to generate new hypotheses[37,78]. Finally new methods are being developed to further leverage the benefits of compendia to improve different types of analyses.

**Challenges integrating across experiments**

While compendia are rich community resources that can be leveraged to gain new insights into transcription, two major challenges make integration across experiments a difficult endeavor: batch effects and strain variation. Batch effects introduced by technical sources (lab that produced the data, sequencing depth) or biological sources (experimental conditions) can either obscure or highlight biological signals, while strain variation can lead to reduced detection of transcription due to incomplete read mapping.

*Batch effects*

In general, batch effects can disrupt detection of biological signal.[97–100] Consequently, it might be expected that compendia, which can integrate many different types of experiments together, require batch correction. However, a recent study by Lee *et al*.[101] examined the effect of technical sources of variability in a compendium setting. They simulated gene expression compendia with varying amounts of technical variability and assessed the ability to detect the original underlying structure in the data after noise was added and then after batch correction was applied. In general, they found that for compendium with a few sources of technical variation batch correction can be effective, however with many more sources of technical variation batch correction isn't necessary and can even start to remove some of the desired biological signal. If correction is applied to a compendium where the experiment-specific noise is largely independent, more of the biological information is removed since biological signals are consistent while noise is experiment specific.

In the case where a compendium contains a few sources of technical variability, like different platforms[54], the dominant signal is the variability between platforms and applying batch correction methods should recover the underlying biological signal. In contrast, in the case where a compendium contains many sources of variability, like many different types of experiments each contributing independent sources of noise, then the aggregation of each experiment-specific source of variability washes out from the underlying biological signal that is consistent across experiments. In this scenario, applying batch correction methods will remove more of the biological signal.

For the cases where batch correction is effective, commonly established methods like Limma[102] and ComBat[103] allow scientists to set sources of variability as covariates.[54] Limma removes technical noise by first fitting a linear model, using *lmFit*, which describes the relationship between the input gene expression and the experimental design labels such as batch assignments and covariates. The resulting model is a coefficient matrix that contains weights for the contribution of the noise component contained in the total observed gene expression matrix. This estimated contribution can be subtracted out from the input expression data. Similarly, Combat also assumes that the input gene expression signal contains an additive batch effect component that can be removed by estimating the batch effect using empirical bayes and subtracting this out.

*Strain variation*

Microbial strain variation further hinders integration across experiments. Strain variation refers to genomic variation that occurs at the sub-species level and can take the form of single nucleotide variants (SNVs) and other small variants, distinct complements of accessory genes, and genomic rearrangements[104]. While strain variation is a critical component of understanding a species' ultimate phenotypic variation, each form of variation causes distinct challenges for integrating expression across strain types. For example, SNVs decrease the average nucleotide identity between the reference sequence used for read quantification and the sample, which can decrease mapping rates non-uniformly across samples.[105] Similarly, the reference sequence may not contain the same set of genes as is present in the sample. This is because most microbial species have a large number of accessory genes, genes which are not universal within that species but are important for endowing unique phenotypic versatility to that strain. Taken together, these accessory genes, which can comprise a significant fraction of the genome (e.g., ~20% for staphylococci[106–108]), comprise a pangenome, defined as the supersets of genes found in the genome of any one species member, including core conserved genes and these accessory genes.[106] When the reference sequence does not contain the same genes as are present in a sample, this can lead to decreased mapping rates and unobserved gene expression.[109] Lasty, genomic rearrangements or insertions may disrupt operon organization for polycistronic transcripts, which may cause difficulties for counting spanning reads that are present in a sample but not represented in a reference.[110] However, integrating strain variation is important not only to understand within-species phenotypic diversity, but also because accessory genes can modify function of the core genome.[111]

Even given these challenges, different approaches have been developed to take advantage of publicly available microbial expression data sets in the face of strain variation. For example, *P. aeruginosa* has five major lineages detected upon genome analyses of over a thousand strains [112] with two major clades that many strains belong to including the widely studied strains PAO1 and PA14[113]. Strains PAO1 and PA14 contain different sets of accessory genes. One common solution is to only consider core genes since they are shared across strain type.[114–118] In order to include accessory genes, separate compendia can be generated so that major strain types (PAO1 and PA14) are separated but there are PAO1-specific genes within the PAO1 compendium.[45]

Most compendia are comprised of a single strain of microorganism (Table 1). This can be achieved by relying on the metadata associated with experiments available in the data repository or using information provided in publications to collect experiments from a single strain. However metadata are notoriously incompletely recorded[119] and difficult to harmonize across studies[120], which may lead to inappropriate inclusion or exclusion of samples in a compendia. Notably, less than half of the publicly available microbial RNA-seq data has been submitted to the Gene Expression Omnibus or Array Express or Expression Atlas. These three platforms provide detailed and standardized meta-data that can be accessed programmatically and easily used in high throughput computational analyses.[121] An alternative approach is to verify the strain annotation using taxonomy assignments provided in the SRA Run Browser analysis tab, or to perform assignment using with a tool like sourmash gather, which selects the minimum set of reference genomes in a database necessary to cover the reads in a sample.[122]

Alternatively, a pangenome could be used as a reference so that core genes are collapsed across strain types while accessory genes are included in the analysis.[109] Using these pangenomes as a reference balances computational cost and fidelity to sample genomes, and can take advantage of databases designed to address similar problems for metagenomic sample processing.[123] This approach was pioneered for the analysis of *Staphylococcus aureus* (*S. aureus*) strains directly from metatranscriptomes, as no reference genome was available with which to perform read quantification. This approach may be successful for building species-wide compendia but needs further research. Indeed, one substantial draw back would be the negation of spanning reads, as pangenomes are typically built from genes and not operons. The increasing use of metatranscriptomics to contextualize a species' function presents computational challenges but also opportunities to identify unique transcriptional signatures in their native and highly complex environment, such as *Haemophilus influenzae* during viral infection, or *S. aureus'* host defense response in the nares.[124,125]

Overall, despite some challenges to constructing compendia, there are existing solutions that make compendia analysis possible and the benefits of the biological discoveries we can glean make it worth it.

**Discussion**

With advancements in high throughput sequencing technology more transcriptome data has become available, presenting opportunities for integration of diverse experiments into compendia. Recent successes of computational methods, especially unsupervised machine learning approaches, have demonstrated that biologically meaningful patterns can be extracted from microbial compendia. Given these recent advances, as well as tools developed in the analysis of human expression compendia, we anticipate development in the computational tool space will continue to drive biological discovery from microbial compendia.

While computational approaches for using heterogeneous compendia have been around for approximately 15 years[76], there remains work to be done to evaluate the computational methods that are most suitable for capturing the transcriptional patterns in compendia. Given the success to date of unsupervised learning methods[64–68,80,81,85,126], and the work that has been done in this space in human expression compendia[127,128], we anticipate that future development and evaluation of these methods will prove useful in the analysis of microbial expression compendia. A comprehensive analysis using human compendia showed that different models and model architectures captured different pathways, revealing that the use of multiple analysis methods led to more complete biological representations.[127] Similarly, there has been some assessment of microbial compendia examining pathway representation using denoising autoencoders[78] and expression changes captured using variational autoencoders[101]. However, an equivalent comprehensive evaluation as undertaken in human compendia is needed to assess the information captured in microbial compendia – what types of signals are captured when the model architecture, regularization, penalty functions, connectivity between layers is varied? This information will determine what model, or range of models, are appropriate for downstream analyses. As new feature extraction models continue to be developed to improve the information captured by and the interpretability of these models, such as through the incorporation of prior information[129], such assessment becomes important to help guide researchers on the computational strategy they use.

Microbial gene expression compendia provide have proven to be a fruitful resource for studying systems-level changes and have been leveraged to infer TRNs[62], make predictions about phenotypes[54], and reveal coordinated gene sets[68,77,78,80]. Furthermore, compendia have been shown to improve the analysis of individual experiments[85]. The advancements in computational tools and webtools, which have made the information in some existing compendia easily accessible, is opening the door to new avenues of research, situating the study of transcription in a global context.

**Acknowledgements:**

This work was supported by grants from the Gordon and Betty Moore Foundation (GBMF4552 to CSG) and Cystic Fibrosis Foundation (HOGAN19GO to DAH and GREENE21GO to CSG).

**Author Contributions:**

**AJL:** Conceptualization; Investigation; Project administration; Writing-original; Writing-review

TR: Conceptualization; Investigation; Writing-original; Writing-review

GD: Writing-review

JO: Writing-review

DAH: Funding acquisition; Writing-review

CSG: Conceptualization; Funding acquisition; Supervision; Writing-review

**Table 1**: Examples of existing microbial compendia.

| Compendium | Organism | Description | No. experiments | No. Samples | No. genes | Platform |
|---|---|---|---|---|---|---|
| *P. aeruginosa* compendium[77] | *P. aeruginosa* | Compendium containing *P. aeruginosa* array data downloaded from ArrayExpress archived in 2014. It includes a mixture of different strain types, media, experimental stimuli. | 109 | 950 | 5,549 | Affymetrix platform GPL84 |
| *P. aeruginosa* RNA-seq compendium[45] | *P. aeruginosa* | Compendium containing *P. aeruginosa* RNA-seq data downloaded from GEO and SRA in 2021. It includes a mixture of different strain types, media, experimental stimuli | > 100 | 2,333 | 5,563 (PAO1) 5,887 (PA14) | RNA-seq |

| Name | Organism | Description | # Conditions | # Genes | # Samples | Platform |
|---|---|---|---|---|---|---|
| EcoMAC[53] | E. coli | Compendium containing E. coli array data downloaded from GEO, ASAP database, ArrayExpress. It includes different strains, media and tests different environmental and genetic perturbations. | 127 | 2,198 | 4,189 | Affymetrix E. Coli Genome 2.0 Array GPL 3154; Affymetrix Ecoli Antisense Array GPL 199 |
| EcoGEC[54] | E. coli | Compendium containing E. coli gene array data from EcoMAC plus RNA-seq data downloaded from GEO. It includes different strains, media and tests different environmental and genetic perturbations. | 144 | 2,262 | 4,166 | Affymetrix E. Coli Genome 2.0 Array; Affymetrix Ecoli Antisense Array; RNA-seq |
| Unnamed[130] | E. coli | Compendium containing E. coli gene array data downloaded from GEO, ArrayExpress and Stanford Microarray Database. It includes a mixture of different experimental conditions | 74 | 870 | NA | Affymetrix; P33; spotted cDNA/DNA; spotted oligonucleotides |
| Unnamed[131,132] | S. cerevisiae | Compendium containing S. cerevisiae array data was a combination of perturbation experiments downloaded from PUMAdb and experiments generated by a genetic screen comparing mutant or compound-treated culture vs wild-type or mock-treated culture. Growth conditions for the screen were consistent across experiment. | >151 | 1,909 | >2000 | two-color cDNA microarray hybridization assay |
| Refine.bio[55] | Many prokaryotes | Database containing processed compendia | ~40 to >500 | ~300 to ~13,000 | ~5000 | microarray; RNA-seq |

| | | for multiple prokaryotes including *P. aerguinosa*, *E. Coli* and *S. cerevisiae*. The data for these compendia were downloaded from SRA, GEO and ArrayExpress. | | | | |
|---|---|---|---|---|---|---|

*Note in SRA, samples are referred to as "Experiment" and a group of samples forming an experiment are referred to as a "Study".